\documentclass[11pt,a4paper,reqno]{amsart}
\usepackage{amsmath}
\usepackage{amsfonts}
\usepackage{amssymb}
\usepackage{amsthm}
\usepackage{newlfont}
\usepackage{a4}
\usepackage{enumerate}
\usepackage[pdftex]{graphicx,color}

\theoremstyle{definition}

\theoremstyle{remark}
\newtheorem{example}{Example}[section]

\usepackage[centertags]{amsmath}
\usepackage{graphicx}
\usepackage{amsfonts}
\usepackage{amssymb}
\usepackage{amsthm}
\usepackage{bbm}
\usepackage{newlfont}
\usepackage{a4}
\usepackage{enumerate}

\newlength{\defbaselineskip}
\newcommand{\setlinespacing}[1]%
           {\setlength{\baselineskip}{#1 \defbaselineskip}}

\newcommand{\map}{\rightarrow}
\newcommand{\q}{\quad}
\renewcommand{\epsilon}{\varepsilon}

\newcommand{\la}{\lambda}

\renewcommand{\rho}{\varrho}
\renewcommand{\phi}{\varphi}

\newcommand{\R}{{\mathbb{R}}}

\newcommand{\N}{{\mathbb N}}

\newcommand{\Com}{{\mathbb C}}

\newcommand{\Z}{\mathbb{Z}}

\newcommand{\setcomb}[2]{
\left\{
\begin{smallmatrix}
#1 \\ #2
\end{smallmatrix}
 \right\}  }

\allowdisplaybreaks

\begin{document}

\title[Two dimensional (anti)symmetric sine functions]
{Two dimensional symmetric and antisymmetric generalizations
 of sine functions}

\author{Ji\v{r}\'{i} Hrivn\'{a}k$^{1,2}$}
\author{Lenka Motlochov\'{a}$^{3}$}
\author{Ji\v{r}\'{i} Patera$^1$}

\date{\today}
\begin{abstract}\ 

Properties of 2-dimensional generalizations of sine functions that are symmetric or antisymmetric with respect to permutation of their two variables are described. It is shown that the functions are orthogonal when integrated over a finite region $F$ of the real Euclidean space, and that they are discretely orthogonal when summed up over a lattice of any density in $F$. Decomposability of the products of functions into their sums is shown by explicitly decomposing products of all types. The formalism is set up for Fourier-like expansions of digital data over 2-dimensional lattices in $F$. Continuous interpolation of digital data is studied.

\end{abstract}\
\maketitle
\noindent
$^1$ Centre de recherches math\'ematiques, Universit\'e de Montr\'eal, C.~P.~6128 -- Centre ville,
Montr\'eal, H3C\,3J7, Qu\'ebec, Canada
\\
$^2$ Department of Physics,  Faculty of Nuclear Sciences and Physical Engineering, Czech
Technical University, B\v{r}ehov\'a~7, 115 19 Prague 1, Czech Republic
\\
$^3$ Department of Mathematics, Faculty of Nuclear Sciences and Physical Engineering, Czech
Technical University, Prague, Trojanova 13, 120 00 Prague 2, Czech Republic

\noindent $\phantom{^\S}$~E-mail: jiri.hrivnak@fjfi.cvut.cz, patera@crm.umontreal.ca

\section{Introduction}

The purpose of this paper is to complete and extend \cite{HP} by considering the remaining two families of special functions and their properties, namely the generalizations of sine functions of two variables, which are either symmetric or antisymmetric with respect to permutations of their two variables. They are denoted here by $\sin^\pm_{(\lambda,\mu)}(x,y),$ where $x,y\in\R$ and $\lambda,\mu\in\N$. The functions are of independent interest. The paper \cite{HP} is devoted to the study of 2-dimensional symmetric and antisymmetric generalizations of the common exponential and cosine functions, namely 
$E^\pm_{(\lambda,\mu)}(x,y)$ and $\cos^\pm_{(\lambda,\mu)}(x,y)$.

In \cite{KPtrig}, the functions denoted $SIN^\pm$ and $COS^\pm$ were introduced for any number of real variables, and their properties were studied. Thus, the functions $\cos^\pm_{(\lambda,\mu)}(x,y)$ of \cite{HP}, as well as the functions $\sin^\pm_{(\lambda,\mu)}(x,y)$ considered here, are the functions of \cite{KPtrig} specialized to $2D$.

Standard trigonometric Fourier decompositions of functions (continuous or discrete) of two variables \cite{Tol} use special functions formed as products of two trigonometric functions, each depending on one variable. The variables are measured along two orthogonal axes. The approach undertaken in \cite{HP}, and extended here, appears to be the only $2D$ `trigonometric' alternative to the standard approach in the literature. Our expansion functions are also built as products of two trigonometric functions, but the two variables are intertwined within each trigonometric function, so that no substitution of variables can bring it to the form used in the standard approach.

Restriction of the functions of \cite{KPtrig} to two variables allows us to be more specific about the details of their properties, most notable being their discretization and orthogonality, continuous and discrete. In particular, analogs of the four types of standard cosine transforms \cite{Strang,mart,Brit}, developed in \cite{HP} and in here, would not be possible without specific description of discrete domains of orthogonality of the expansion functions applicable in the four cases.

Decomposition of all products of pairs of functions into their sums is described here. It was not considered elsewhere. There are three types of products of sine functions, $\sin^+\sin^+,$ \ $\sin^+\sin^-$, and  $\sin^-\sin^-$, with the same arguments $x,y$. For completeness, we also decompose all products $\sin^\pm\cos^\pm$, and $\cos^\pm\cos^\pm$. The general structure of decompositions of products is rather interesting. It is summarized symbolically in Table~\ref{products}.

\begin{table}[h]
\footnotesize
\addtolength{\tabcolsep}{-3pt}

\begin{center}
\begin{tabular}{|c||c|c|c|c|c|c|c|c|c|c|}
\hline
product &$\sin^+\sin^+$  
        &$\sin^+\sin^-$   
        &$\sin^-\sin^-$
        &$\sin^+\cos^+$
        &$\sin^+\cos^-$
        &$\sin^-\cos^+$
        &$\sin^-\cos^-$
        &$\cos^+\cos^+$
        &$\cos^+\cos^-$ 
        &$\cos^-\cos^-$  \\
\hline 
terms   &$\cos^+$ 
        &$\cos^-$
        &$\cos^+$
        &$\sin^+$
        &$\sin^-$
        &$\sin^-$
        &$\sin^+$
        &$\cos^+$ 
        &$\cos^-$ 
        &$\cos^+$   \\
\hline

\end{tabular}
\bigskip
\caption{Structure of the ten types of products decomposed in the paper. The second row shows the functions appearing in all  the terms of a decomposition.} \label{products}
\end{center}
\end{table}

There are numerous reasons motivating the study of $\sin^\pm$ and $\cos^\pm$ functions in more than one variable. One such reason is the ever presence of trigonometric functions in applications of mathematics from the elementary to the most sophisticated. Our immediate motivation is to be able to use the functions in Fourier-like analysis and interpolation \cite{Davis} of digital data in $2D$.

We describe four versions of the sine transforms of functions given on lattices. They differ by certain small shifts of their arguments which allow orthogonality of the expansion functions to be maintained. These transforms 
correspond to the well known cosine transforms, labeled as $I$, $II$, $III$, and $IV$, where the expansion functions have separated variables \cite{Strang}, namely $\cos(\pi mx)\cdot\cos(\pi ny)$. The four variants of the antisymmetric sine transforms and the four variants of the symmetric sine transforms correspond to different boundary conditions at the boundaries of the fundamental domain. Similarly, there are different types of ordinary cosine and sine tranforms in $1D$ \cite{Brit}.   

In Section~2 we consider the sine functions, followed by some general properties of sine and cosine functions in Section~3. Discretization of the antisymmetric and symmetric sine transforms are the subject of Section~4 together with their interpolations of types I, II, III, and IV. Three remarks are found in the last Section.

\section{Continuous sine transforms in $\R^2$}

First we consider the antisymmetric, then symmetric $2D$ sine transforms. The discretization of transforms is described in Section~4. Here the definitions of symmetric and antisymmetric functions $\cos^\pm(x,y)$ are recalled because they appear in decompositions of products of sine functions into their sums.

\subsection{Antisymmetric sine functions}

\subsubsection{Definitions, symmetries and general properties}\

\begin{figure}[!ht]
\resizebox{2.4cm}{!}{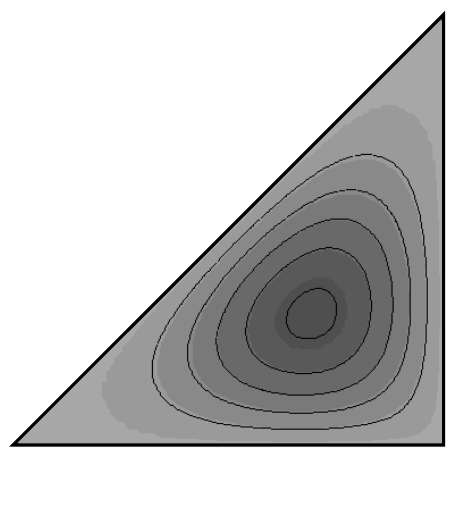}\hspace{22pt}
\resizebox{2.4cm}{!}{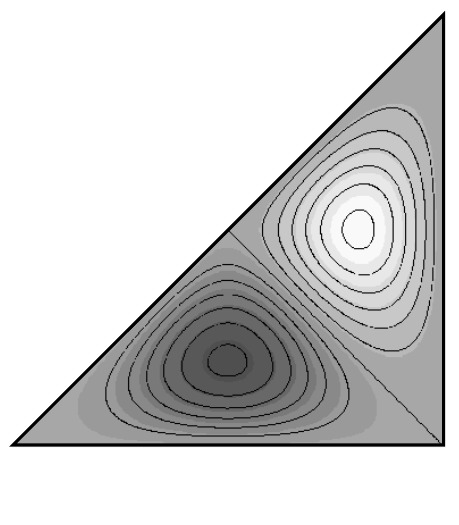}\hspace{22pt}
\resizebox{2.4cm}{!}{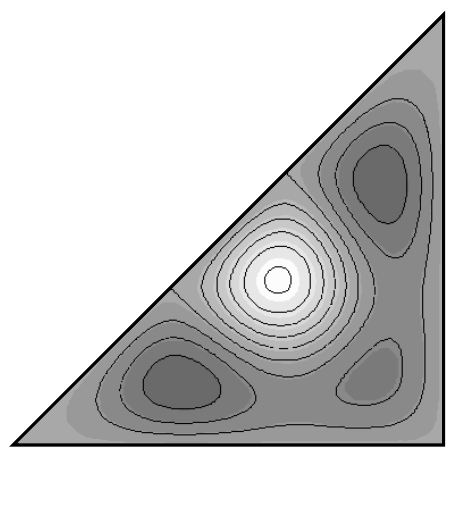}
\\\vspace{2pt}
\resizebox{2.4cm}{!}{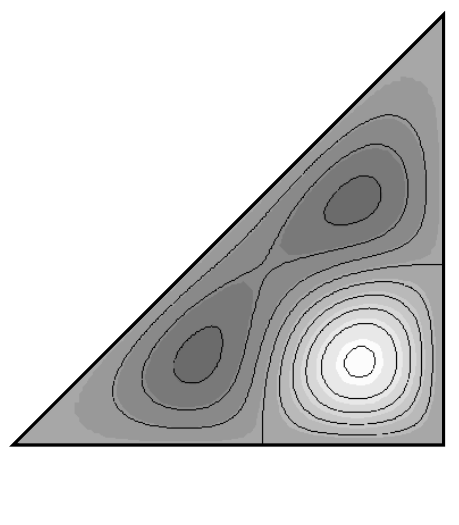}\hspace{22pt}
\resizebox{2.4cm}{!}{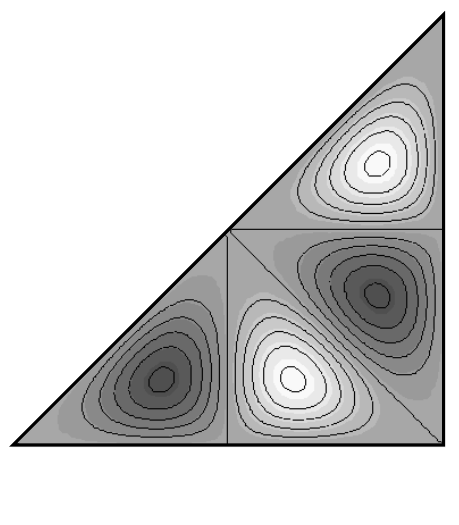}\hspace{22pt}
\resizebox{2.4cm}{!}{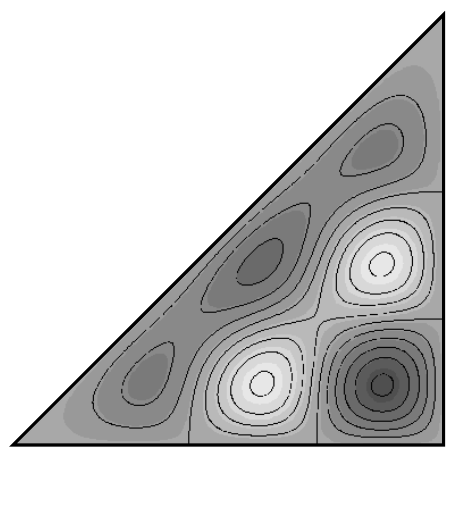} \\ 

\caption{The contour plots of examples of functions $\sin^-_{(k,l)}(x,y)$ in the fundamental domain.}\label{FSm}
\end{figure}

The antisymmetric sine functions \ $\sin^-_{(\lambda,\mu)}(x,y):\ \R^2\map \R$ \ are defined as follows:
\begin{equation}\label{sin-def}
\sin^-_{(\lambda,\mu)}(x,y)
     =\left|\begin{smallmatrix}
     \sin(\pi\lambda x)&\sin(\pi\lambda y)\\
     \sin(\pi\mu x)&\sin(\pi\mu y)\\
     \end{smallmatrix}\right|^-
     =\sin(\pi\lambda x)\sin(\pi\mu y)
     -\sin(\pi\mu x)\sin(\pi\lambda y),\qquad\lambda,\mu,x,y\in\R\,.
\end{equation}
Clearly the functions are continuous and have continuous derivatives of all degrees in $\R^2$. A few examples of functions are shown in Figure~\ref{FSm}. Note that instead of the factor $2\pi$, which was used in \cite{KPtrig}, we use the 'half' argument $\pi$.

The following properties of the functions are verified directly from the definition \eqref{sin-def}:
\begin{align}
\sin^-_{(\lambda,\lambda)}(x,y)&=0 \label{indexvanish}\\
\sin^-_{(\lambda,0)}(x,y)      &= \sin^-_{(0,\mu)}(x,y)=0 \label{indexzeros} \\
\sin^-_{(\lambda,\mu)}(x,y)    &=-\sin^-_{(\mu,\lambda)}(x,y)\label{indexpermut}\\
\sin^-_{(-\lambda,\mu)}(x,y)   &= \sin^-_{(\lambda,-\mu)}(x,y)
       =-\sin^-_{(-\lambda,-\mu)}(x,y)=-\sin^-_{(\lambda,\mu)}(x,y) \label{indexsign}\\
\sin^-_{(\lambda,\mu)}(x,x)    &=0\label{vanish}\\
\sin^-_{(\lambda,\mu)}(x,0)    &= \sin^-_{(\lambda,\mu)}(0,y)=0\\
\sin^-_{(\lambda,\mu)}(x,y)    &=-\sin^-_{(\la,\mu)}(y,x)\label{permut}\\
\sin^-_{(\lambda,\mu)}(-x,y)   &= \sin^-_{(\lambda,\mu)}(x,-y)
               =-\sin^-_{(\lambda,\mu)}(-x,-y)=-\sin^-_{(\lambda,\mu)}(x,y)\label{sign}.
\end{align}
Because of \eqref{indexvanish} -- \eqref{indexsign}, we consider only $\sin^-_{(\lambda,\mu)}$ with $\la>\mu>0$.

In addition, the functions $\sin^-_{(k,l)}$ with $k,l\in\Z$ have symmetries related to the periodicity of the sine function
\begin{equation}\label{asinper}
    \sin^-_{(k,l)}(x+2r,y+2s)= \sin^-_{(k,l)}(x,y),\q r,s\in \Z
\end{equation}
and  
\begin{equation}\label{asinnul}
    \sin^-_{(k,l)}(r,s)=0,\q r,s\in \Z. 
\end{equation}

The relations \eqref{permut} -- \eqref{asinper} imply that it is sufficient to consider the functions 
$\sin^-_{(k,l)}(x,y$) on the closed triangle $F(S_2^{\mathrm{aff}})$ given by its vertices in $\R^2$:
\begin{equation}\label{fund}
(x,y)\in F(S_2^{\mathrm{aff}})= \{(0,0),\,(1,0),\,(1,1)\},
\end{equation}
called the fundamental domain of the affine symmetric group \cite{KPexp}.
The relations \eqref{vanish} and \eqref{asinnul} imply that $\sin^-_{(k,l)}(x,y)$ vanishes on the boundary $\partial F(S_2^{\mathrm{aff}})$  of the fundamental domain.

The graphs of a few lowest functions $\sin^-_{(k,l)}(x,y)$ in the fundamental domain are plotted in Figure~\ref{FSm}.

\subsubsection{Continuous orthogonality}\

The functions $\sin^-_{(k,l)}$ are pairwise orthogonal on the fundamental domain $F(S_2^{\mathrm{aff}})$, 
\begin{equation}
    \int_{F(S_2^{\mathrm{aff}})} \sin^-_{(k,l)}(x,y)\sin^-_{(k',l')}(x,y)\,dx\, dy=\frac{1}{4}\delta_{kk'}\delta_{ll'}, \q k,l,k',l'\in\N,\ k>l,k'>l'
\end{equation}
Any function $f:\R^2\map \Com$ that is antisymmetric $f(x,y)=-f(y,x)$, periodic $f(x+2r,y+2s)= f(x,y),\, r,s\in \Z$, has continuous derivatives and vanishes on $\partial F(S_2^{\mathrm{aff}})$ can be expanded in the antisymmetric sine functions $\sin^-_{(k,l)}$: 
\begin{equation}
f(x,y)=\sum_{\setcomb{k,l\in \N}{k>l}} c_{kl} \sin^-_{(k,l)}(x,y)
,\hspace{1.5cm}
c_{kl} = 4 \int_{F(S^{\mathrm{aff}}_2)} f(x,y)
\sin^-_{(k,l)}(x,y)\,dx\, dy.
 \end{equation}

\subsection{Symmetric sine functions}\ 

\subsubsection{Definitions, symmetries and general properties}\

\begin{figure}[!ht]
\resizebox{2.4cm}{!}{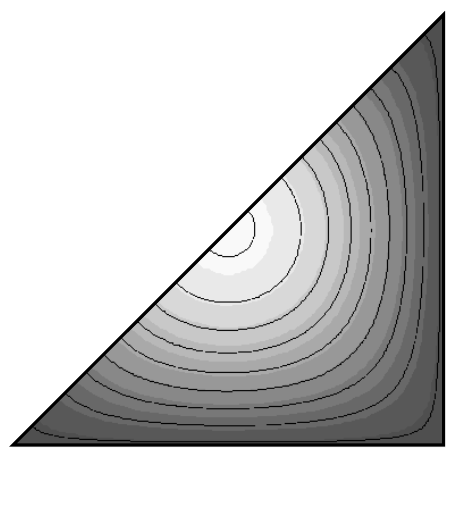}\hspace{22pt}
\resizebox{2.4cm}{!}{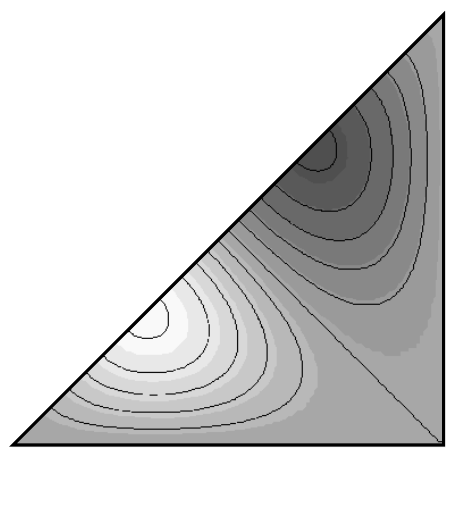}\hspace{22pt}
\resizebox{2.4cm}{!}{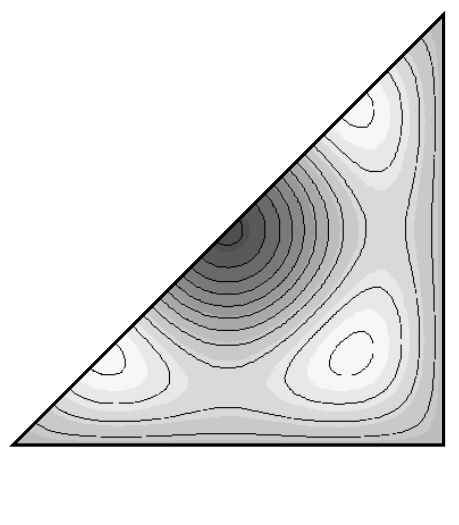}
\\\vspace{2pt}
\resizebox{2.4cm}{!}{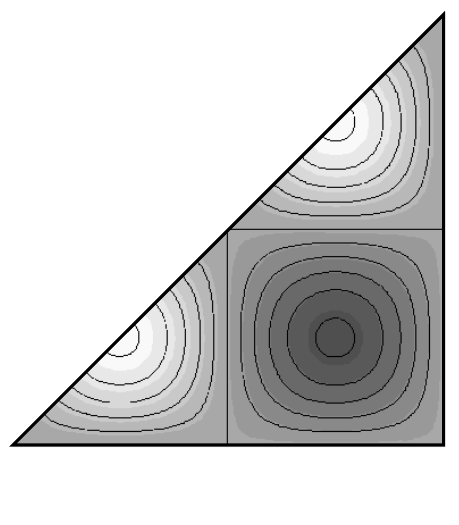}\hspace{22pt}
\resizebox{2.4cm}{!}{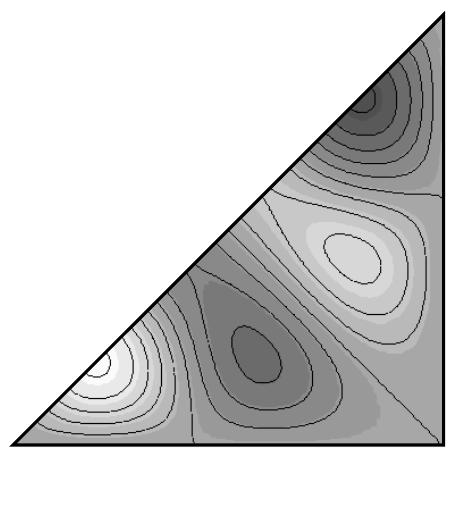}\hspace{22pt}
\resizebox{2.4cm}{!}{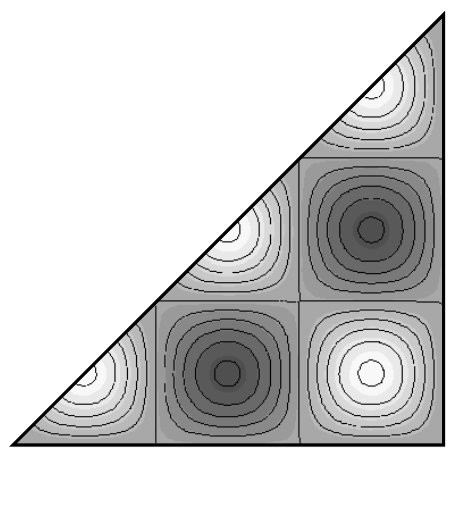} \\ 

\caption{The contour plots of examples of functions $\sin^+_{(k,l)}(x,y)$ in the fundamental domain.}\label{FSp}
\end{figure}

Two-dimensional symmetric sine functions $\sin^+_{(\lambda,\mu)}:\R^2\map \Com$ are defined for $\lambda,\mu\in \R$ in the following form:
\begin{equation}\label{sin+def}
\sin^+_{(\lambda,\mu)}(x,y)
     =\left|\begin{smallmatrix}
     \sin(\pi\lambda x)&\sin(\pi\lambda y)\\
     \sin(\pi\mu x)&\sin(\pi\mu y)\\
     \end{smallmatrix}\right|^+
     =\sin(\pi\lambda x)\sin(\pi\mu y)
     +\sin(\pi\mu x)\sin(\pi\lambda y)\\
\end{equation}
A few examples of functions are shown in Figure~\ref{FSp}. Note that instead of the factor $2\pi$ used in \cite{KPtrig}, we use the 'half' argument $\pi$.

The following properties of the functions are verified directly from the definition \eqref{sin+def}:
\begin{align}
 \sin^+_{(\lambda,0)}(x,y) &= \sin^+_{(0,\mu)}(x,y)=0\label{sindexvanish}\\
 \sin^+_{(\lambda,\mu)}(x,y)&=\sin^+_{(\mu,\lambda)}(x,y)\label{sindexpermut}\\
 \sin^+_{(-\lambda,\mu)}(y,x)&=\sin^+_{(\lambda,-\mu)}(x,y)
        =-\sin^+_{(-\lambda,-\mu)}(x,y)=-\sin^+_{(\lambda,\mu)}(x,y)\label{sindexsign}\\
\sin^+_{(\lambda,\mu)}(x,0)&=\sin^+_{(\lambda,\mu)}(0,y)=0 \\
\sin^+_{(\lambda,\mu)}(x,y)&=\sin^+_{(\lambda,\mu)}(y,x),\label{ssinpermut}\\
\sin^+_{(\lambda,\mu)}(-x,y)&=\sin^+_{(\lambda,\mu)}(x,-y)
         =-\sin^+_{(\lambda,\mu)}(-x,-y)=-\sin^+_{(\lambda,\mu)}(x,y).\label{ssinsign}
\end{align}

Because of \eqref{sindexvanish} -- \eqref{sindexsign}, we consider only such $\sin^+_{(\lambda,\mu)}$ with $\lambda\geq\mu>0$.

The functions $\sin^+_{(k,l)}$ with $k,l \in \Z$ have symmetries related to the periodicity of sine function:
\begin{align}
    \sin^+_{(k,l)}(x+2r,y+2s)&= \sin^+_{(k,l)}(x,y),\q r,s\in \Z,\label{ssinper}\\
    \sin^+_{(k,l)}(r,s)&=0,\q r,s \in\Z.
\end{align}

The relations (\ref{ssinpermut}) -- (\ref{ssinper}) imply that it is sufficient to consider the functions $\sin^+_{(k,l)},\ k,l\in\N,\ k\geq l$ on the fundamental domain $F(S_2^{\mathrm{aff}})$ \cite{KPexp}.

The graphs of the lowest symmetric sine functions $\sin^+_{(k,l)}$, $k,l\in\{1,\dots,3\},\,k\geq l$ in the fundamental domain are plotted in Figure~\ref{FSp}.

\subsubsection{Continuous orthogonality}\

The functions $\sin^+_{(k,l)}$ are mutually orthogonal on the fundamental domain $F(S_2^{\mathrm{aff}})$,
\begin{equation}
    \int_{F(S_2^{\mathrm{aff}})} \sin^+_{(k,l)}(x,y)
\sin^+_{(k',l')}(x,y)\,dx\, dy=
\frac{G_{kl}}{4}\delta_{kk'}\delta_{ll'}, \q k,l,k',l' \in\N,\, k\geq l,k' \geq l',  \end{equation}
where $G_{kl}$ is defined by
\begin{equation}
	G_{kl}=\begin{cases}2&k=l\\1&\textrm{otherwise}\end{cases}.
\end{equation}
Any function $f:\R^2\map \Com$ that is symmetric $f(x,y)=f(y,x)$, periodic $f(x+2r,y+2s)= f(x,y),\, r,s\in \Z$, has continuous derivatives and vanishes for $x=1$ and $y=0$ can be expanded in the symmetric sine functions $\sin^+_{(k,l)}$:
\begin{equation}
f(x,y)=\sum_{\setcomb{k,l\in \N}{k\geq l}} c_{kl} \sin^+_{(k,l)}(x,y)
,\hspace{1.5cm}
c_{kl} = 4G_{kl}^{-1} \int_{F(S^{\mathrm{aff}}_2)} f(x,y)
\sin^+_{(k,l)}(x,y)\,dx\, dy.
 \end{equation}

\section{Additional properties of $2D$ trigonometric functions.}

Properties of $2D$ generalizations of sine and cosine are closely interwoven. Let us first recall definitions of the functions $\cos^\pm_{(\lambda,\mu)}(x,y)$ according to \cite{HP,KPtrig}.
\begin{equation}\label{cosdef}
\cos^\pm_{(\lambda,\mu)}(x,y)
     =\left|\begin{smallmatrix}
     \cos(\pi\lambda x)& \cos(\pi\lambda y)\\
     \cos(\pi\mu x)& \cos(\pi\mu y)\\
     \end{smallmatrix}\right|^\pm
     = \cos(\pi\lambda x) \cos(\pi\mu y)
     \pm \cos(\pi\mu x) \cos(\pi\lambda y)\,,\\
\end{equation}
where the upper or lower signs should be taken simultaneously.

\subsection{Laplace and other differential operators}\ 

An obvious relation between $2D$ sine and cosine functions arises from the second derivatives,
\begin{gather}\label{derivative}
\frac{\partial^2}{\partial x \partial y}\sin^\pm_{(\la,\mu)}(x,y)=\pi^2\la\mu\cos^\pm_{(\la,\mu)}(x,y)\,,\qquad
\frac{\partial^2}{\partial x \partial y}\cos^\pm_{(\la,\mu)}(x,y)=\pi^2\la\mu\sin^\pm_{(\la,\mu)}(x,y)\,.
\end{gather}

The functions $\sin^\pm_{(\lambda,\mu)}(x,y)$ are eigenfunctions of the Laplace operator
\begin{equation*}
\left(\frac{\partial^2}{\partial x^2}+\frac{\partial^2}{\partial y^2}\right)\sin^\pm_{(\la,\mu)}(x,y)=-\pi^2(\la^2+\mu^2)\sin^\pm_{(\la,\mu)}(x,y)
\end{equation*}
as well as of the operator \eqref{derivative} applied twice:
\begin{equation*}
\frac{\partial^2}{\partial x^2}\frac{\partial^2}{\partial y^2}\sin^\pm_{(\la,\mu)}(x,y)=\pi^4\la^2\mu^2\sin^\pm_{(\la,\mu)}(x,y).
\end{equation*}

Functions $\sin^+$ satisfy the equality
\begin{equation*}
\frac{\partial}{\partial \mathrm{\textbf{n}}}\sin^+_{(\la,\mu)}(x,x)=0, 
\end{equation*}
where \textbf{n} is normal to the boundary $x=y$.

\subsection{Product decompositions}\ 

Products of two $\sin^\pm_{(\lambda,\mu)}(x,y)$ functions decompose into the sum of cosine functions. Products of $\cos^\pm_{(\lambda,\mu)}(x,y)$ functions decompose into the sum of cosine functions, while mixed products,
$\sin^\pm_{(\lambda,\mu)}(x,y)\cdot\cos^\pm_{(\lambda',\mu')}(x,y)$, decompose into the sum of sine functions, see Table~\ref{products}. 
Common trigonometrical identities,  
\begin{equation}\label{identity}
\begin{aligned}
  &2\sin(pz)\sin(qz)=-\cos((p+q)z)+\cos((p-q)z),\\ 
  &2\cos(pz)\cos(qz)= \cos((p+q)z)+\cos((p-q)z),\\
  &2\sin(pz)\cos(qz)= \sin((p+q)z)+\sin((p-q)z),
\end{aligned}
\end{equation}
are used when decompositions of products of $\sin^\pm_{(\lambda,\mu)}(x,y)$ and $\cos^\pm_{(\lambda,\mu)}(x,y)$ are calculated.

There are altogether ten different types of products to consider. A concise presentation of the ten decompositions is found in Table~\ref{decomposition}. The following examples are intended to illustrate how the actual decomposition is obtained from the Table.

\begin{table}
\begin{center}

\setlength{\tabcolsep}{1ex}
\begin{tabular}{|c|c|c@{\kern.5ex}c@{\kern.5ex}c@{\kern.5ex}c|c@{\kern.5ex}c@{\kern.5ex}c@{\kern.5ex}c|}
\hline
\multicolumn{2}{|c|}{}& \multicolumn{4}{c|}{$(\la\pm\la',\mu\pm\mu')$} & \multicolumn{4}{c|}{$(\la\pm\mu',\mu\pm\la')$}\\
\cline{3-10}
\multicolumn{2}{|c|}{}& $(+,+)$& $(+,-)$& $(-,+)$& $(-,-)$& $(+,+)$& $(+,-)$& $(-,+)$&$(-,-)$\\
\hline

$\sin^+\cdot\sin^+$& $\cos^+$& $+$&$-$&$-$&$+$&$+$&$-$&$-$&$+$\\ \hline
$\sin^+\cdot\sin^-$& $\cos^-$& $+$&$-$&$-$&$+$&$-$&$+$&$+$&$-$\\ \hline
$\sin^-\cdot\sin^-$& $\cos^+$& $+$&$-$&$-$&$+$&$-$&$+$&$+$&$-$\\ \hline
$\sin^+\cdot\cos^+$& $\sin^+$& $+$&$+$&$+$&$+$&$+$&$+$&$+$&$+$\\ \hline
$\sin^+\cdot\cos^-$& $\sin^-$& $+$&$+$&$+$&$+$&$-$&$-$&$-$&$-$\\ \hline
$\sin^-\cdot\cos^+$& $\sin^-$& $+$&$+$&$+$&$+$&$+$&$+$&$+$&$+$\\ \hline
$\sin^-\cdot\cos^-$& $\sin^+$& $+$&$+$&$+$&$+$&$-$&$-$&$-$&$-$\\ \hline
$\cos^+\cdot\cos^+$& $\cos^+$& $+$&$+$&$+$&$+$&$+$&$+$&$+$&$+$\\ \hline
$\cos^+\cdot\cos^-$& $\cos^-$& $+$&$+$&$+$&$+$&$-$&$-$&$-$&$-$\\ \hline
$\cos^-\cdot\cos^-$& $\cos^+$& $+$&$+$&$+$&$+$&$-$&$-$&$-$&$-$\\ \hline
\end{tabular}

\medskip

\caption{Decomposition of all products of functions $\sin^\pm(x,y)$ and $\cos^\pm(x,y)$ into the sum of eight such functions. The first column shows the product. The second column contains the function appearing in the decomposition terms. Subsequent columns provide: (i) the subscripts labeling the decomposition terms (first line); (ii) the pair of signs applicable in the subscripts of each term; (iii) the signs in front of each term in the decomposition (remaining lines).}\label{decomposition}
\end{center}
\end{table}

\begin{example}\ 

The top left entry of the table identifies the line where the decomposition of products of the form $\sin^+_{(\lambda,\mu)}(x,y)\cdot\sin^+_{(\lambda',\mu')}(x,y)$ is given. For our example, we selected $(\la,\mu)=(2,1)$, $(\la',\mu')=(3,2)$. The second column entry on the same line indicates that all terms of the decomposition are the functions $\cos^+_{(\alpha,\beta)}(x,y)$. The subsequent eight entries on that line refer to individual terms of the decomposition. Each term has a coefficient $\pm1$. The actual sign of a term is shown on the line. The value of $\alpha$ and $\beta$ is identified on the two top lines of the column. Thus for the first term of the example we find at the top line that $\alpha=2\pm3$ and $\beta=1\pm2$. Which of the two signs is applicable in $\alpha$ and $\beta$ is specified at the head of the column of the first term as $(+,+)$, so that $\cos^+_{(\alpha,\beta)}(x,y)=\cos^+_{(5,3)}$. Subsequent terms are identified in a similar way. The symbol $(x,y)$ is omitted to simplify the expression. From \eqref{identity} follows the presence of the coefficient 4 multiplying each product below.

In this way, some of the subscripts of the terms of the decomposition are negative. A convenient convention is to label the terms by non-negative subscripts. Therefore the symmetries of the functions need to be used in order to write the subscripts as positive. If a subscript of a term $\sin^\pm_{(\alpha,\beta)}$ should be 0, the sine term vanishes.

The four specific examples illustrate the decompositions:

\begin{align*}
&4\sin^+_{(2,1)}\cdot\sin^+_{(3,2)}
     =\cos^+_{(5, 3)}
     -\cos^+_{(5,-1)}
     - \cos^+_{(-1,3)}
     + \cos^+_{(-1,-1)}
     + \cos^+_{(4,4)}
     - \cos^+_{(4,-2)}
     -\cos^+_{(0,4)}
     + \cos^+_{(0,-2)}\\
&\hspace{2.7cm}
     =\cos^+_{(5, 3)}
     -\cos^+_{(5,1)}
     - \cos^+_{(3,1)}
     + \cos^+_{(1,1)}
     + \cos^+_{(4,4)}
     - \cos^+_{(4,2)}
     -\cos^+_{(4,0)}
     + \cos^+_{(2,0)}
\end{align*}
\begin{align*}
&4\sin^+_{(2,1)}\cdot\sin^-_{(3,2)}
    =\cos^-_{(5, 3)}
     -\cos^-_{(5,-1)}
     - \cos^-_{(-1, 3)}
     + \cos^-_{(-1,-1)}
     - \cos^-_{(4,4)}
     + \cos^-_{(4,-2)}
     +\cos^-_{(0,4)}
     - \cos^-_{(0,-2)}\\
&\hspace{2.7cm}
     =\cos^-_{(5, 3)}
     -\cos^-_{(5,1)}
     + \cos^-_{(3, 1)}
     + 0
     - 0
     + \cos^-_{(4,2)}
     -\cos^-_{(4,0)}
     + \cos^-_{(2,0)}
\end{align*}
\begin{align*}
&4\sin^+_{(2,1)}\cdot\cos^+_{(3,2)}
    =\sin^+_{(5, 3)}
     + \sin^+_{(5,-1)}
     +\sin^+_{(-1,3)}
     + \sin^+_{(-1,-1)}
     +\sin^+_{(4,4)}
     + \sin^+_{(4,-2)}
     + \sin^+_{(0,4)}
     + \sin^+_{(0,-2)}\\
&\hspace{2.7cm}
    =\sin^+_{(5, 3)}
     - \sin^+_{(5,1)}
     -\sin^+_{(3,1)}
     + \sin^+_{(1,1)}
     +\sin^+_{(4,4)}
     - \sin^+_{(4,2)}
     + 0
     + 0
\end{align*}
\begin{align*}
&4\sin^-_{(2,1)}\cdot\cos^+_{(3,2)}
    =\sin^-_{(5, 3)}
     + \sin^-_{(5,-1)}
     +\sin^-_{(-1,3)}
     + \sin^-_{(-1,-1)}
     +\sin^-_{(4,4)}
     + \sin^-_{(4,-2)}
     + \sin^-_{(0,4)}
     + \sin^-_{(0,-2)}\\
&\hspace{2.7cm}
     =\sin^-_{(5, 3)}
     - \sin^-_{(5,1)}
     +\sin^-_{(3,1)}
     + 0
     + 0
     - \sin^-_{(4,2)}
     + 0
     + 0   
\end{align*}
\end{example}

\section{Discrete transforms}

The four versions of the discrete sine transforms introduced here use a grid of points $(x_m,y_n)$~\cite{HP}, extending over finite regions that differ by their boundaries. The grid is defined by the numbers $N$, $T$, and $b$. The number of points of the grid is $N^2$, constant $b\in [0,1]$ is the displacement of lattice points from their original position,  and $T$ determines argument of sine functions, so that the argument is equal to $\frac{2\pi}{T}$. To make the distinctions between the four versions of the antisymmetric and symmetric transforms easy to compare, we first describe the pertinent regions.

Define the `closed square' $K_{[a,a']}$ by
\begin{equation}
K_{[a,a']}:=[a,a']\times [a,a'],
\qquad\qquad a,a'\in\Z\,.
\end{equation}

For the antisymmetric discrete sine transforms, it is necessary to define a 'partly-open triangle' $K^-_{[0,1]}$: 
\begin{equation}
K_{[0,1]}^-:=\big\{(x,y) \in [0,1]\times [0,1]\mid x>y\big\}.
\end{equation} 
Similarly, for the symmetric sine transforms, we introduce a `closed triangle' $K^+_{[0,1]}$, which contains one additional side, described by $x=y$, in comparison with the triangle for antisymmetric sine transforms:
\begin{equation}
K_{[0,1]}^+:=\left\{(x,y)\in [0,1]\times [0,1]\mid x\geq y \right\}.
\end{equation}

\subsection{Antisymmetric discrete sine transforms}\

The antisymmetric sine functions are closely related to the antisymmetric exponential functions~\cite{HP}. Four types of discrete antisymmetric sine transforms can be derived from the antisymmetric exponential transforms. The idea is to suitably extend given functions and then apply the antisymmetric exponential transforms from~\cite{HP}. 
 
In order to derive discrete antisymmetric sine transforms, we define the following three functional extension operators. They extend a complex function defined on the `partly-opened triangle' $K^-_{[0,1]}$ to functions defined on the `closed square' $K_{[-L,L]}$.

First, let $f:K_{[0,L]}\map\Com$. We define its extension $E_L f:K_{[-L,L]}\map\Com$  
\begin{equation}\label{extE}
E_Lf(x,y):=\begin{cases} f(x,y)  &\hspace{0.3cm} 0\leq x\leq L, \hspace{0.3cm} 0<y<L \\ 
                       -f(-x,y)  & -L\leq x<0,  \hspace{0.3cm}0\leq y\leq L \\ 
                       -f(x,-y)  &\hspace{0.3cm} 0\leq x\leq L, -L\leq y<0 \\ 
                       f(-x,-y)  & -L\leq x<0, -L\leq y<0,\\ 
            \end{cases}
\end{equation}

Secondly, let $f:K_{[0,1]}\map\Com$. We define its extension $Rf:\mathrm K_{[0,2]}\map\Com$ to the square as follows
\begin{equation}\label{extR}
Rf(x,y):=\begin{cases}  
                      f(x,y)     & 0 \leq x \leq 1,\ 0 \leq y \leq 1 \\ 
                      f(2-x,y)   & 1 < x \leq 2,   \ 0 \leq y \leq 1 \\ 
                      f(x,2-y)   & 0 \leq x \leq 1,\ 1< y \leq 2 \\ 
                      f(2-x,2-y) & 1 < x \leq 2,   \ 1 < y \leq 2 .
         \end{cases}
\end{equation}
For the function $f:K^-_{[0,1]}\map\Com$, we define antisymmetric extensions $Af:K_{[0,1]}\map\Com$ as
\begin{equation}\label{aext}
 Af(x,y)=\begin{cases}
f(x,y) & x>y\\
0 & x=y \\
-f(y,x) & x<y.
\end{cases}
\end{equation}

In the first two types of discrete antisymmetric transforms, we consider the given functions $f_1:K^-_{[0,1]}\map\Com$ that have zero values for the lines $x=1$ and $y=0$. In the last two types, we consider the given functions $f_2:K^-_{[0,1]}\map\Com$ that vanish on the line $y=0$. The additional conditions on the boundary are derived from properties of sine functions on the regions in view.

The following extensions of a function $f_1$ or $f_2$ with corresponding values of $N,T$ and $b$ are substituted into the formula (56) from~\cite{HP}:

\begin{alignat}{7}
(I)  \quad&E_1Af_1  &&:\quad K_{[-1,1]} &&\map \Com, \qquad\text{ where }&\quad N&=2M,\quad &T&=2,&\quad b&=1  \\
(II) \quad&E_1Af_1  &&:\quad K_{[-1,1]} &&\map \Com, \qquad\text{ where }&\quad N&=2M,\quad &T&=2,&\quad b&=1/2\\
(III)\quad&E_2RAf_2 &&:\quad K_{[-2,2]} &&\map \Com, \qquad\text{ where }&\quad N&=4M,\quad &T&=4,&\quad b&=1  \\
(IV) \quad&E_2RAf_2 &&:\quad K_{[-2,2]} &&\map \Com, \qquad\text{ where }&\quad N&=4M,\quad &T&=4,&\quad b&=1/2
\end{alignat}

We notice that $E_1Af_1$ has zero values on the boundary of $K_{[-1,1]}$, on the axes $x,y$ and on the lines $x=\pm y$, and $E_2RAf_2$ vanishes on the boundary of $K_{[-2,2]}$, on the axes $x,y$ and on the lines $x=\pm y$. The different conditions of zero values are due to the fact that in the first and second type of antisymmetric sine transforms, we consider $\sin^-_{(k,l)}(x,y)$ in comparison with the third and fourth type, where we consider $\sin^-_{(k,l)}(\frac{x}{2},\frac{x}{2})$. 

Due to (anti)symmetry and zero values of the extended functions on axes $x,y$, on the lines $x=\pm y$ and on borders of $K_{[-1,1]}$ or $K_{[-2,2]}$, we obtain the final explicit form of the four antisymmetric sine transforms:

\begin{enumerate}[{AMDST-} I.]
\bigskip

 \item  
 $$
 \psi^{\mathrm{I},-}_M(x,y)=\sum_{\setcomb{k,l=1}{k>l}}^{M-1} c_{k,l}^{\mathrm{I},-}\sin^-_{(k,l)}(x,y),
\qquad
 c_{k,l}^{\mathrm{I},-}=\frac{4}{M^2} \sum_{\setcomb{m,n=1}{m>n}}^{M-1}f_1\left(x_m,y_n\right)\sin^-_{(k,l)}\left(x_m,y_n\right)
 $$
where $x_m=\frac{m}{M}$, $y_n=\frac{n}{M}$.
\bigskip\bigskip
 
\item
$$
\psi^{\mathrm{II},-}_M(x,y)=\sum_{\setcomb{k,l=1}{k>l}}^{M} c_{k,l}^{\mathrm{II},-}\sin^-_{(k,l)}(x,y),\qquad c_{k,l}^{\mathrm{II},-}=\frac{4d_{k,M} d_{l,M}}{M^2} \sum_{\setcomb{m,n=0}{m>n}}^{M-1}f_1\left(x_m,y_n\right)\sin^-_{(k,l)}\left(x_m,y_n\right)
$$
where $x_m=\frac{m+\frac12}{M}$, $y_n=\frac{n+\frac12}{M}$,\quad $d_{M,M}=\tfrac12$ and 
$d_{k,M}=1$ for $k\neq M$.
\bigskip\bigskip

\item  
\begin{align*}
\psi^{\mathrm{III},-}_M(x,y)&=\sum_{\setcomb{k,l&=0}{k>l}}^{M-1} c_{k,l}^{\mathrm{III},-}\sin^-_{(k+\frac12,l+\frac12)}(x,y)\,,\\
 c_{k,l}^{\mathrm{III},-}&=\frac{4}{M^2} \sum_{\setcomb{m,n=1}{m>n}}^{M}d_{m,M} d_{n,M}f_2\left(x_m,y_n\right)\sin^-_{(k+\frac12,l+\frac12)}\left(x_m,y_n\right) 
\end{align*}
where $x_m=\frac{m}{M}$, $y_n=\frac{n}{M}$.
\bigskip\bigskip

\item  
$$
\psi^{\mathrm{IV},-}_M(x,y)=\sum_{\setcomb{k,l=0}{k>l}}^{M-1} c_{kl}^{\mathrm{IV},-}\sin^-_{(k+\frac12,l+\frac12)}(x,y),\qquad 
c_{kl}^{\mathrm{IV},-}=\frac{4}{M^2} \sum_{\setcomb{m,n=0}{m>n}}^{M-1}f_2\left(x_m,y_n\right)\sin^-_{(k+\frac12,l+\frac12)}\left(x_m,y_n\right)
$$
where $x_m=\frac{m+\frac12}{M}$, $y_n=\tfrac{n+\tfrac12}{M}$
\end{enumerate}

\bigskip

\subsubsection{Example of antisymmetric sine interpolation}\ 

We proceed as follows: Choose a continuous model function, sample its value on a lattice, develop the digital data into the finite series according to one of the transforms above, interpolate the digital data, and compare the resulting continuous function with the model function.

\begin{figure}[!ht]
\resizebox{3.2cm}{!}{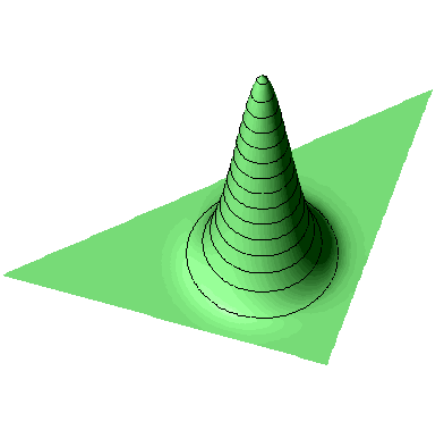}\hspace{1.2cm}
\resizebox{2.4cm}{!}{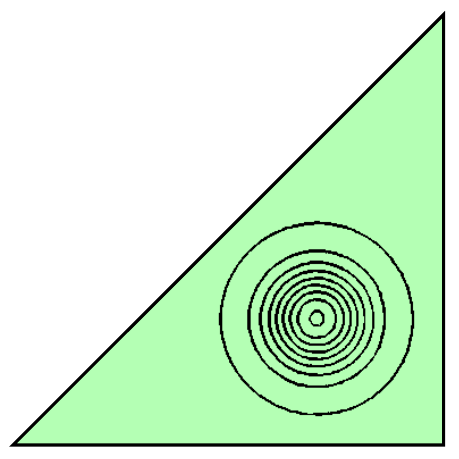}

\caption{The function $f$ of~(\ref{f4}) is plotted over the domain $F(S_2^{\mathrm{aff}})$.}\label{Ff}
\end{figure}

Our model function is the Gaussian distribution
\begin{equation}\label{f4}
 f(x,y)=e^{- \frac{(x-x')^2+(y-y')^2}{2\sigma^2}}\,,
\end{equation}
with the following parameters $(x',y')=(0.707,\,0.293)$ and $\sigma=0.079$. The function $f$, restricted to the domain $F(S_2^{\mathrm{aff}})$, is shown in Figure~\ref{Ff}.

\begin{figure}[!ht]
\resizebox{2.4cm}{!}{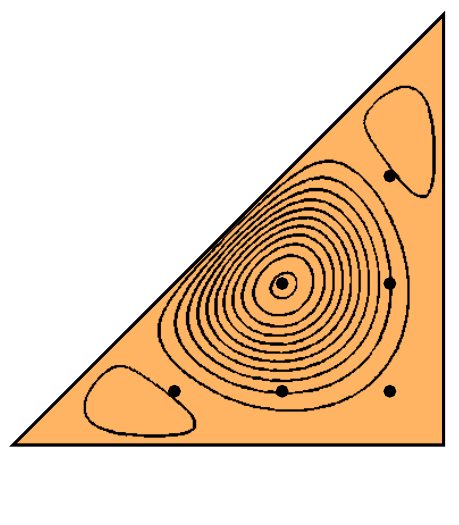}\hspace{1.3cm}
\resizebox{2.4cm}{!}{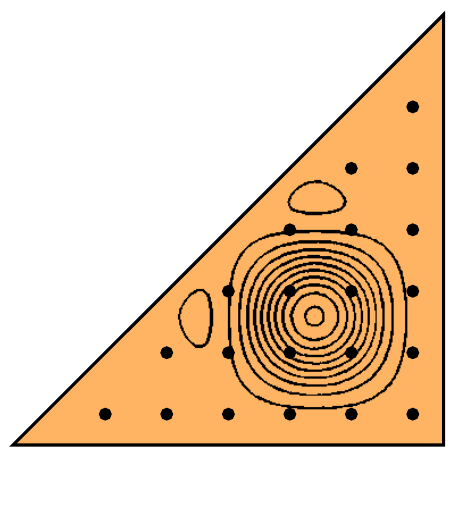}\hspace{1.3cm}
\resizebox{2.4cm}{!}{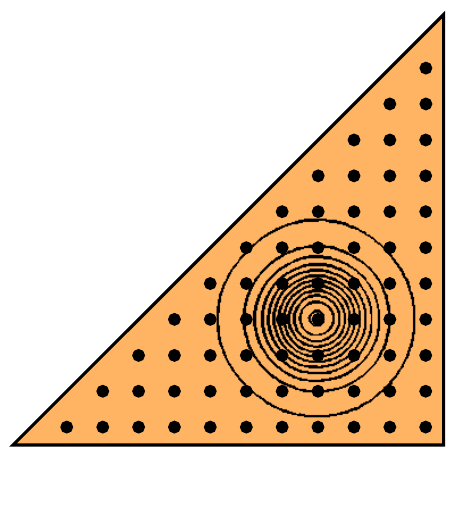}
\\\vspace{1pt}
\resizebox{3.2cm}{!}{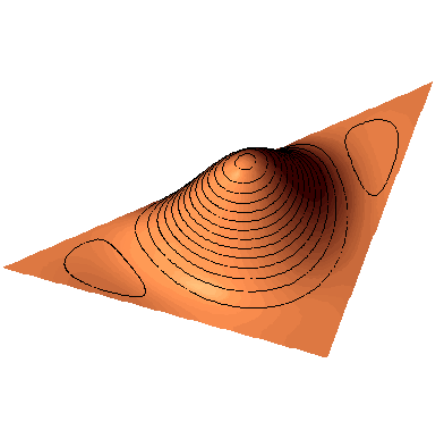}\hspace{14pt}
\resizebox{3.2cm}{!}{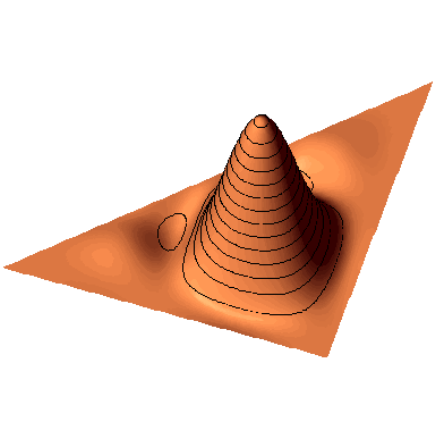}\hspace{14pt}
\resizebox{3.2cm}{!}{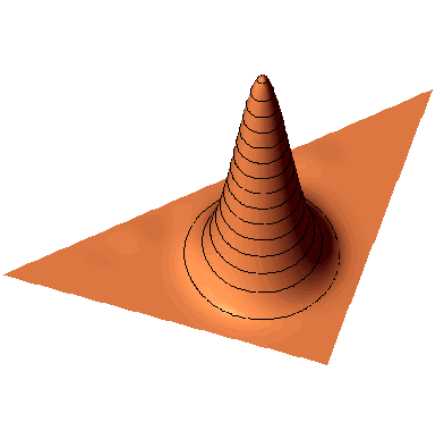}
\\\vspace{1pt}
\resizebox{3.2cm}{!}{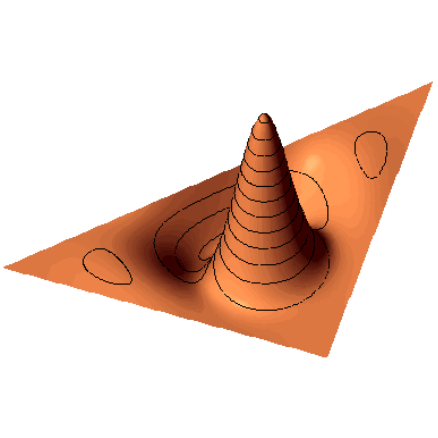}\hspace{14pt}
\resizebox{3.2cm}{!}{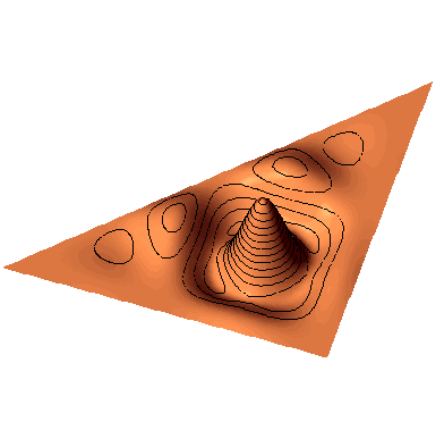}\hspace{14pt}
\resizebox{3.2cm}{!}{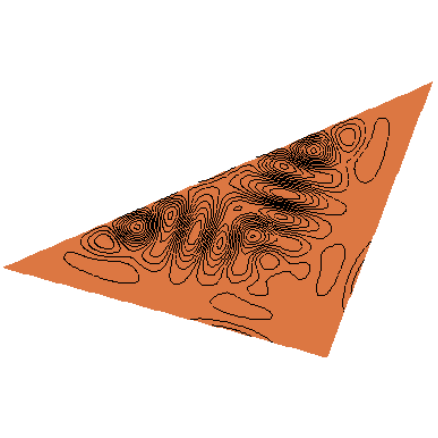}
\caption{Three examples \ 
$\psi^{\mathrm{II},-}_4$, $\psi^{\mathrm{II},-}_7$, and $\psi^{\mathrm{II},-}_{12}$ of antisymmetric sine functions of the type AMDST-II interpolating the function $f$ in Figure~\ref{Ff}. Sampling points for the interpolation are shown as small black dots. Figures in the third line depict the difference $f-\psi^{\mathrm{II},-}_N$.}\label{FfCm}
\end{figure}

We calculate the antisymmetric interpolating sine functions of the type AMDST-II, namely $\psi^{\mathrm{II},-}_4$, $\psi^{\mathrm{II},-}_7$ and $\psi^{\mathrm{II},-}_{12}$, using grids of density given by $M=4$, $7$, and $12$. The interpolating functions, together with the sampling grids, are presented in Figure~\ref{FfCm}. 

\subsection{Symmetric discrete sine transforms}\

Symmetric sine functions are related to symmetric exponential functions \cite{HP}. Four types of discrete symmetric sine functions are derived from the four types of symmetric exponential transforms. We apply symmetric exponential transforms from \cite{HP} to the extensions of a given function.

In order to derive discrete symmetric sine transforms, we use functional operators $E_L$, $R$, and for the functions $f:K^+_{[0,1]}\map\Com$, we define symmetric extension $Sf:K_{[0,1]}\map\Com$ by the formula   
\begin{equation}
 Sf(x,y)=\begin{cases}
f(x,y) & x\geq y\\
f(y,x) & x<y.
\end{cases}
\end{equation}

In the first two types of discrete symmetric transforms, we consider the given functions $f_1:K^+_{[0,1]}\map\Com$ that have zero values for the lines $x=1$ and $y=0$. In the last two types, we consider the given functions $f_2:K^+_{[0,1]}\map\Com$ that vanish on the line $y=0$. The additional conditions on the boundary are derived from properties of sine functions on the regions in view.

The following extension of a function $f_1$ or $f_2$ with the corresponding values of $N,T$ and $b$ are substituted into the formula (65) from \cite{HP}:
\begin{alignat}{7}
(I)  \quad&E_1Sf_1  &&:\quad K_{[-1,1]} &&\map \Com, \qquad\text{ where }&\quad N&=2M,\quad 
                     &T&=2,&\quad b&=1  \\
(II) \quad&E_1Sf_1  &&:\quad K_{[-1,1]} &&\map \Com, \qquad\text{ where }&\quad N&=2M,\quad 
                     &T&=2,&\quad b&=1/2  \\
(III)\quad&E_2Sf_2  &&:\quad K_{[-2,2]} &&\map \Com, \qquad\text{ where }&\quad N&=4M,\quad 
                     &T&=4,&\quad b&=1  \\
(IV) \quad&E_2Sf_2  &&:\quad K_{[-2,2]} &&\map \Com, \qquad\text{ where }&\quad N&=4M,\quad 
                     &T&=4,&\quad b&=1/2  
\end{alignat}

We notice that $E_1Sf_1$ has zero values on the boundary of $K_{[-1,1]}$ and on the axes $x,y$, and $E_2RSf_2$ vanishes on the boundary of $K_{[-2,2]}$ and on the axes $x,y$. The different conditions of zero values are due to the fact that in the first and second type of symmetric sine transforms, we consider $\sin^-_{(k,l)}(x,y)$ in comparison with the third and fourth type, where we consider $\sin^-_{(k,l)}(\frac{x}{2},\frac{x}{2})$. 

Due to (anti)symmetry and zero values of the extended functions on the axes $x,y$ and on borders of $K_{[-1,1]}$ or $K_{[-2,2]}$, we obtain the final explicit form of the four symmetric sine transforms:

\begin{enumerate}[{SMDST--}I.]
 \item  $$\psi^{\mathrm{I},+}_M(x,y)=\sum_{\setcomb{k,l=1}{k\geq l}}^{M-1} c_{kl}^{\mathrm{I},+}\sin^+_{(k,l)}(x,y),$$ $$ c_{kl}^{\mathrm{I},+}=\frac{4}{M^2G_{kl}} \sum_{\setcomb{m,n=1}{m\geq n}}^{M-1}G_{mn}^{-1}f_1\left(x_m,y_n\right)\sin^+_{(k,l)}\left(x_m,y_n\right)$$
where $x_m=\frac{m}{M}$, $y_n=\frac{n}{M}$.

 \item  $$\psi^{\mathrm{II},+}_M(x,y)=\sum_{\setcomb{k,l=1}{k\geq l}}^{M} c_{k,l}^{\mathrm{II},+}\sin^+_{(k,l)}(x,y),\q c_{k,l}^{\mathrm{II},+}=\frac{4}{M^2G_{kl}} \sum_{\setcomb{m,n=0}{m\geq n}}^{M-1}G_{mn}^{-1}f_1\left(x_m,y_n\right)\sin^+_{(k,l)}\left(x_m,y_n\right)$$
where $x_m=\frac{m+\frac12}{M}$, $y_n=\frac{n+\frac12}{M}$.

 \item  $$\psi^{\mathrm{III},+}_M(x,y)=\sum_{\setcomb{k,l=0}{k\geq l}}^{M-1} c_{k,l}^{\mathrm{III},+}\sin^+_{(k+\frac12,l+\frac12)}(x,y),$$ $$  c_{k,l}^{\mathrm{III},+}=\frac{4}{M^2G_{kl}} \sum_{\setcomb{m,n=1}{m\geq n}}^{M}d_{m,M}d_{n,M}G_{mn}^{-1}f_2\left(x_m,y_n\right)\sin^+_{(k+\frac12,l+\frac12)}\left(x_m,y_n\right) $$
where $x_m=\frac{m}{M}$, $y_n=\frac{n}{M}$.
 \item  $$\psi^{\mathrm{IV},+}_M(x,y)=\sum_{\setcomb{k,l=0}{k\geq l}}^{M-1} c_{kl}^{\mathrm{IV},+}\sin^+_{(k+\frac12,l+\frac12)}(x,y),$$ $$ c_{kl}^{\mathrm{IV},+}=\frac{4}{M^2G_{kl}} \sum_{\setcomb{m,n=0}{m\geq n}}^{M-1}G_{mn}^{-1}f_2\left(x_m,y_n\right)\sin^+_{(k+\frac12,l+\frac12)}\left(x_m,y_n\right)$$
where $x_m=\frac{m+\frac12}{M}$, $y_n=\frac{n+\frac12}{M}$.
\end{enumerate}

\subsubsection{Example of symmetric sine interpolation}\

\begin{figure}[!ht]
\resizebox{2.4cm}{!}{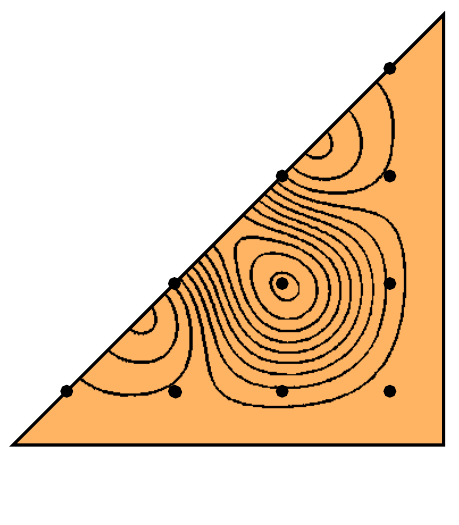}\hspace{1.3cm}
\resizebox{2.4cm}{!}{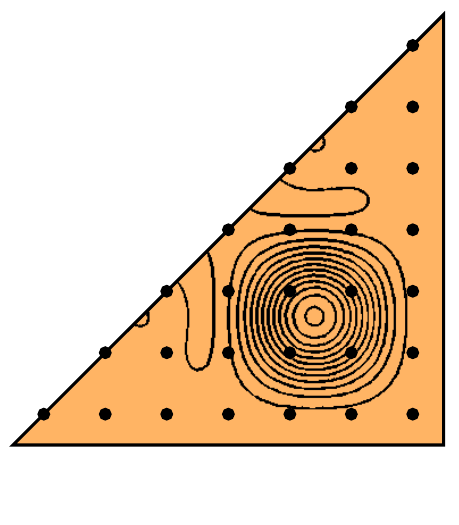}\hspace{1.3cm}
\resizebox{2.4cm}{!}{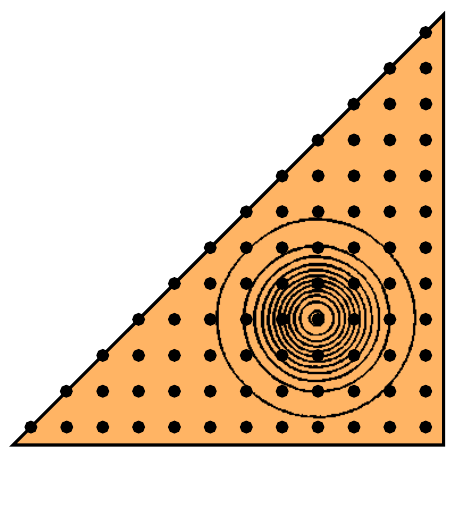}
\\\vspace{1pt}
\resizebox{3.2cm}{!}{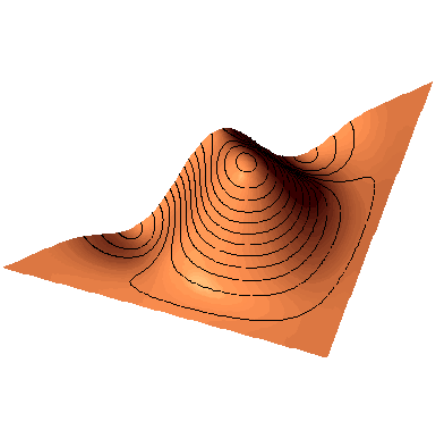}\hspace{14pt}
\resizebox{3.2cm}{!}{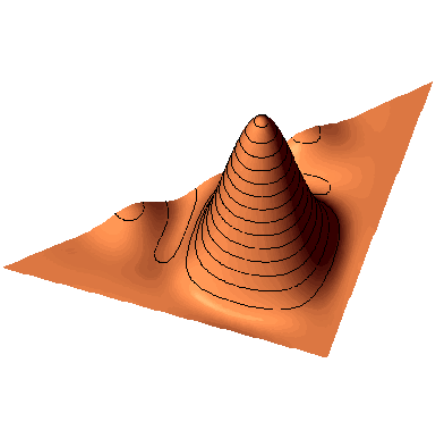}\hspace{14pt}
\resizebox{3.2cm}{!}{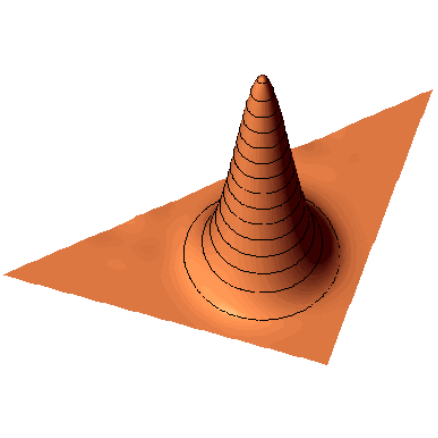}
\\\vspace{1pt}
\resizebox{3.2cm}{!}{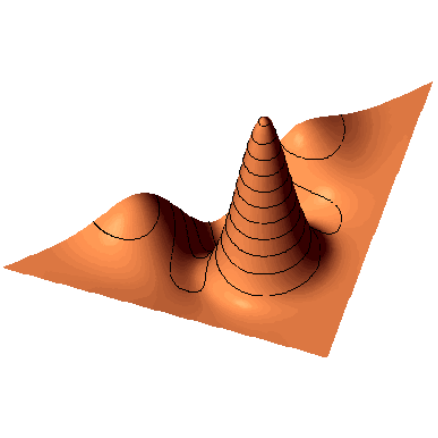}\hspace{14pt}
\resizebox{3.2cm}{!}{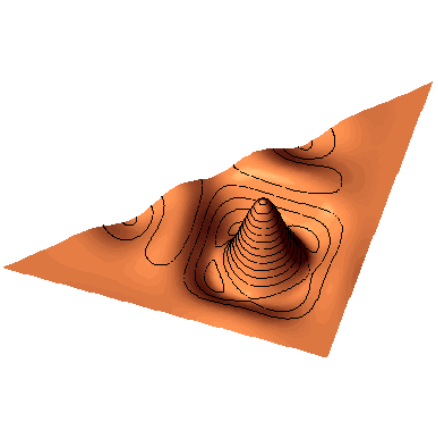}\hspace{14pt}
\resizebox{3.2cm}{!}{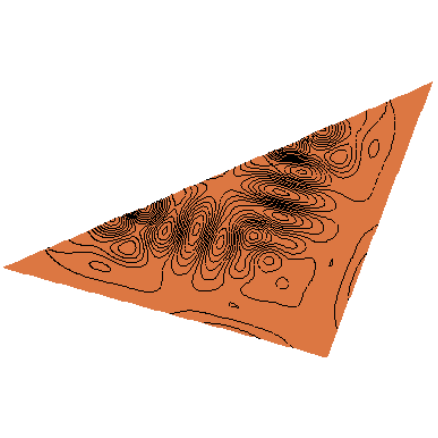}
\caption{Three examples \ 
$\psi^{\mathrm{II},+}_4$, $\psi^{\mathrm{II},+}_7$ and $\psi^{\mathrm{II},+}_{12}$ of symmetric sine functions of the type SMDST-II interpolating the function $f$ in Figure~\ref{Ff}. Sampling points for the interpolation are shown as small black dots. The third line  show the difference $f-\psi^{\mathrm{II},+}_N$.}\label{FfSp}
\end{figure}

Consider the Gaussian distribution $f$ defined by (\ref{f4}). We calculate the symmetric sine interpolating functions of the type SMDST-II: $\psi^{\mathrm{II},+}_4$, $\psi^{\mathrm{II},+}_7$ and $\psi^{\mathrm{II},+}_{12}$. These interpolating functions, together with the interpolating grids, are depicted in Figure~\ref{FfSp}. 

\section{Concluding remarks}

Decomposition of products of functions described in the article for all pairs of functions $\sin^\pm_{(\lambda,\mu)}$ and $\cos^\pm_{(\lambda,\mu)}$ paves the way to a wealth of new properties of these functions \cite{MP}, such as a large variety of trigonometric-like identities, representation of functions as orthogonal polynomials, and recursion relations for their construction.
 \smallskip
 
In the paper, we make no use of a particular property of the four families of functions, which is proving useful elsewhere \cite{MP}: The functions of each family split into two mutually exclusive congruence classes according to the value of the sum of their subscript $\lambda+\mu\mod2$. For example, it can be seen in Table~\ref{decomposition} that all terms in the decomposition of one product belong to the same congruence class, and that the classes add up during multiplication of functions.
   \smallskip

The well known Weyl formula for the character of an irreducible finite dimensional representation of semisimple Lie groups is a ratio of two $S$-functions studied in \cite{KP2}. The character functions are endowed with many properties that are fundamental to the theory of representations in general. The $\sin^\pm$ functions in the paper, which depend on two variables, resemble the $S$-functions of the group $SU(3)$. In fact, in the case of one variable, these functions coincide. It is therefore interesting to explore properties of functions formed as an analogous ratio of $\sin^\pm$ functions.
 \smallskip

\section*{Acknowledgements}

Work supported by the Natural Sciences and Engineering Research Council of Canada, and in part also by MITACS, and by the MIND Research Institute. J.~H. acknowledges the postdoctoral fellowship, and with L.~M. they are grateful for the hospitality extended to them at the Centre de recherches math\'ematiques, Universit\'e de Montr\'eal, where most of the work was carried out.


\end{document}